\documentclass[twocolumn,showpacs,showkeys,amsmath,amssymb]{revtex4-1}

\usepackage[dvips]{graphicx}
\usepackage{dcolumn}
\usepackage{bm}

  \newcommand{\nn}{\nonumber}

\begin{document}

\title{Anomalous dimension of the gauge invariant canonical decomposition for proton momentum with the background field method}

\author{Yoshio Kitadono}
 \email{kitadono@impcas.ac.cn}

\author{Pengming Zhang}
 \email{zhpm@impcas.ac.cn}
 \affiliation{Institute of Modern Physics, Chinese Academy of Sciences,\\
Lanzhou, People's Republic of China, 730000.}

\date{\today}

\begin{abstract}
The anomalous dimension for the gauge-invariant-canonical decomposition of the energy-momentum tensor for quarks and gluons is studied by the background field method. In particular, the consistency between the background field method and the renormalization in the gluonic sectors is investigated. The analysis shows that the naive gauge-invariant-decomposition has an inconsistency between its definition and the renormalization in the background field method. Although we try to consider a trick to overcome this inconsistency in computing the anomalous dimension, the gauge-parameter dependence remains in the final result. This result should be extended to the problems on the gauge-invariant-canonical-spin decomposition.   
\end{abstract}


\maketitle

One of remained problems in quantum chromodynamics (QCD) is the complete decomposition of the proton spin into the orbital angular momentum and the spin in terms of quarks and gluons. Although several decompositions are known \cite{JM_decomp,Ji_decomp}, the complete decomposition of the proton spin with keeping gauge invariance to each term seemed impossible. However, in the past few-years, a new kind of the gauge-invariant decomposition was proposed by Chen {\sl et al.} \cite{Chen1,Chen2}. A prediction based on their decomposition is that the momentum fraction carried by gluons in the proton is about one-fifth \cite{Chen2}; this is different from the well-known value, about half, predicted by the standard QCD \cite{gamma_QCD}. Their study triggered many debates and caused much controversy (see Refs.~\cite{Leader_Review, Wakamatsu_Review} for recent reviews).

After Chen {\sl et al.} proposed the new decomposition, Wakamatsu extended their splitting-technique to the following covariant-form \cite{Wakamatsu_cova},
\begin{equation}
 A^{\mu} = A^{\mu}_{\rm \small pure} + A^{\mu}_{\rm \small phys},
\end{equation}
with the condition $F^{\mu\nu}_{\rm \small pure} = 0$. 
The field $A^{\mu}_{\rm \small pure}$ carries the degree of freedom of the gauge transformation and the field $A^{\mu}_{\rm \small phys}$ carries the physical degree of freedom and $F^{\mu\nu}_{\rm \small pure}$ is the field strength for $A_{\rm \small pure}$.   
After long debates, two types of decompositions are widely accepted as the gauge-invariant decomopositions, namely, gauge-invariant-canonical (gic) and the gauge-invariant-kinetic (gik) decompositions. Although originally this decomposition was for the three-dimensional operators, currently four-dimensionally-covariant decompositions both for the generalized-angular-momentum tensor and the energy-momentum tensor are known \cite{Leader_Review, Wakamatsu_Review}.
   
The gic decomposition of the energy-momentum tensor $T^{\mu\nu}_{\rm \small gic}$ is given by
\begin{eqnarray}
 T^{\mu\nu}_{\rm \small gic} &=& T^{\mu\nu}_{\rm \small gic,q} +  T^{\mu\nu}_{\rm \small gic,g},\nn\\
  T^{\mu\nu}_{\rm \small gic,q} &=& \frac{1}{2}\bar{\psi}\gamma^{\{\mu}iD^{\nu\}}_{\rm \small pure}\psi,\nn\\
  T^{\mu\nu}_{\rm \small gic,g} &=& -\mbox{Tr}\left[F^{\{\mu\alpha}D^{\nu\}}_{\rm \small pure}A_{\alpha,\rm\small phys}\right], \label{eq.gic.Tmunu}
\end{eqnarray}
where $a^{\{\mu}b^{\nu\}}=a^{\mu}b^{\nu}+a^{\nu}b^{\mu}$ is the symmetrization symbol, $D^{\mu}_{\rm \small pure} = \partial^{\mu}-ig[A^{\mu}_{\rm\small pure},~~~]$ is the covariant derivative with $A^{\mu}_{\rm\small pure}$, and we ignored the terms including $g^{\mu\nu}$ since these terms are irrelevant to the momentum operators of quarks and gluons. On the other hand, Belinfante-improved-energy-momentum tensor $T^{\mu\nu}_{\rm \small Bel}$ \cite{Belinfante_Rosenfeld} as a gik decomposition is given by
\begin{eqnarray}
  T^{\mu\nu}_{\rm \small Bel} &=& T^{\mu\nu}_{\rm \small Bel,q} +  T^{\mu\nu}_{\rm \small Bel,g},\nn\\
  T^{\mu\nu}_{\rm \small Bel,q} &=& \frac{1}{2}\bar{\psi}\gamma^{\{\mu}iD^{\nu\}}\psi,\nn\\
  T^{\mu\nu}_{\rm \small Bel,g} &=& -\mbox{Tr}\left[F^{\{\mu\alpha}F^{\nu\}}{}_{\alpha}\right], \label{eq.Bel.Tmunu}
\end{eqnarray}
where these two-definitions are related to each other by the surface term, $T^{\mu\nu}_{\rm\small gic}=T^{\mu\nu}_{\rm\small Bel}-\partial_{\alpha}\mbox{Tr}\left(F^{\{\mu\alpha}A^{\nu\}}_{\rm\small phys}\right)$. The quark (gluon) momentum $\vec{P}_{q(g)}$ is related to the energy-momentum tensor $T^{\mu\nu}_{q(g)}$ through the relation $P^{i}_{q(g)}=\int d^3x T^{0i}_{q(g)}$. The gauge transformations for split fields are given in
\begin{eqnarray}
 A^{\mu~\prime}_{\rm \small pure}(x) &
 =& U(x)\left( A^{\mu}_{\rm \small pure}(x) + \frac{i}{g}\partial^{\mu} \right)U^{\dagger}(x),\nn\\
   A^{\mu~\prime}_{\rm \small phys}(x) &
 =& U(x)A^{\mu}_{\rm \small phys}(x) U^{\dagger}(x), \label{eq.gauge.trans.law}
\end{eqnarray}
where $U(x)=e^{igt^a\theta^a(x)}$ is the gauge-transformation factor. The two definitions in Eqs.~(\ref{eq.gic.Tmunu}) and (\ref{eq.Bel.Tmunu}) are invariant under the above gauge-transformations.

The gauge transformations in Eq.~(\ref{eq.gauge.trans.law}) are well known in the background field method (BFM). The BFM is an alternative method to quantize a field theory with keeping the gauge invariance manifestly by integrating out the quantum part of the gauge field; it gives the same result with the standard quantization (see the review \cite{Abbott_BFM_Review}). Actually, such a similarity between the gauge transformation in Eq.~(\ref{eq.gauge.trans.law}) and the BFM has already been pointed out by Lorc{\'e} in the context of a path dependence of Wilson lines \cite{Lorce_Path} and Noether theorem under the presence of the background field \cite{Lorce_Noether}, and by Zhang and Pak in the context of the gluon's helicity \cite{Zhang_Pak}. However, the one-loop analysis of the gic decomposition by the BFM has not yet carried out. In this Letter, we push forward to study the recent progress on the problems of the momentum (spin) decompositions by the BFM. In particular, we investigate the anomalous-dimension matrix in Ref.~\cite{Chen1} again by the BFM, because the gic decomposition in Eq.~(\ref{eq.gic.Tmunu}) obviously includes the split gauge-fields and hence one needs the BFM to handle these definitions.

The momentum fraction carried by quarks and gluons in the proton can be predicted by solving the following renormalization-group-equation,
\begin{eqnarray}
\frac{d}{d\ln\mu}
 \left(
   \begin{array}{c}
    \vec{P}_{q}(\mu) \\
    \vec{P}_{g}(\mu) \\
   \end{array}
  \right)
  &=& - \gamma
   \left(
   \begin{array}{c}
    \vec{P}_{q}(\mu) \\
    \vec{P}_{g}(\mu) \\
   \end{array}
  \right), 
\end{eqnarray}
where the mixing matrix $\gamma$ is called the anomalous-dimension matrix and the matrix elements are defined by
\begin{eqnarray}
\gamma &\equiv& - \frac{\alpha_s(\mu)}{4\pi}
\left(
 \begin{array}{cc}
   \gamma^{qq} &  \gamma^{qg}  \\
   \gamma^{gq} &  \gamma^{gg}
\end{array}
\right),
\end{eqnarray}
with the conventional overall-sign to compare ours with recent results. Chen {\sl et al.}'s result of the anomalous dimension \cite{Chen2} is summarized in the following form:
\begin{eqnarray}
\gamma_{\rm \tiny Chen}
&=&  - \frac{\alpha_s(\mu)}{4\pi}
\left(
 \begin{array}{cc}
   - \frac{2}{9}n_g &   \frac{4}{3}n_f \\
     \frac{2}{9}n_g & - \frac{4}{3}n_f
\end{array}
\right),\label{eq.gamma.Chen}
\end{eqnarray}
where $n_g=8$ is the number of gluons and $n_f$ is a number of quark flavors. This anomalous dimension gives the asymptotic limit to gluon momentum, $\vec{P}_{g} = \frac{n_g}{n_g + 6n_f} \vec{P}_{q+g}$; it predicts that the momentum fraction carried by gluons is about one-fifth for $n_f=5$.
The above mixing-matrix is different from the standard-QCD result \cite{gamma_QCD},
\begin{eqnarray}
\gamma_{\rm \tiny QCD}
&=& - \frac{\alpha_s(\mu)}{4\pi}
\left(
 \begin{array}{cc}
   - \frac{8}{9}n_g &   \frac{4}{3}n_f \\
     \frac{8}{9}n_g & - \frac{4}{3}n_f
\end{array}
\right), \label{eq.gamma.QCD}
\end{eqnarray}
where the above standard-results gives the asymptotic limit, $\vec{P}_{g} = \frac{2n_g}{2n_g + 3n_f} \vec{P}_{q+g}$; it predicts that the momentum fraction carried by gluons is about half for $n_f=5$.

The origin of the contradicted result of the asymptotic limit for the momentum fraction carried by the gluons in the proton is the anomalous dimension in Eq.~(\ref{eq.gamma.Chen}). Actually, the evaluation of the anomalous dimension based on Eq.~(\ref{eq.gic.Tmunu}), independently of Ref.~\cite{Chen2}, is studied in Ref.~\cite{Wakamatsu_mom_gamma} and the author concluded that Eq.~(\ref{eq.gic.Tmunu}) leads to the well known result in Eq.~(\ref{eq.gamma.QCD}). However, the author adopted the own method introducing a projector to the physical field $A_{\rm \small phys}$ \cite{Wakamatsu_Feynmanrule}, not the BFM.

In the BFM, one decomposes the gauge field $A^{\mu}$ into the {\sl background field or classical field} $A^{\mu}_{\rm \small bg}$ and the {\sl quantum field} $A^{\mu}_{\rm \small qt}$ as $A^{\mu}=A^{\mu}_{\rm \small bg}+A^{\mu}_{\rm \small qt}$. 
To evaluate the anomalous dimension based on the BFM at the one-loop order, one has to take into account a difference between these two-fields; however, this difference never appears in the standard quantization, namely, two different-roles of the background field $A^{\mu}_{\rm \small bg}$ and quantum field $A^{\mu}_{\rm \small qt}$. The quantum field $A^{\mu}_{\rm \small qt}$ should be integrated out in the sense of the path integral and hence it only describes quantum effects. On the other hand, the background field $A^{\mu}_{\rm \small bg}$ receives that quantum effects as the feedback and it describes the gauge-invariant-effective theory. In other words, the quantum field should be understood as internal propagators and the background field should be understood as external lines in loop diagrams. Therefore, to identify $A^{\mu}_{\rm \small pure}$ as the background field and $A^{\mu}_{\rm \small phys}$ as the quantum field in the BFM, we have to keep the following principle:
 \begin{eqnarray}
 \mbox{background~field:}~A_{\rm \small pure}
  &=& \mbox{external~lines},\nn\\
  \mbox{quantum~field:}~A_{\rm \small phys}
   &=& \mbox{internal~lines}.
\end{eqnarray}
The above principle shows that one cannot introduce the field renormalization to $A_{\rm \small phys}$ in the effective theory after the integration of $A_{\rm \small phys}$, but one should introduce it to $A_{\rm \small pure}$. If one does not correctly treat these differences, the gauge invariance of the effective theory after the loop integral is not guaranteed by the definition of the BFM \cite{Abbott_BFM_Review}.

\begin{figure}[htb]
  \begin{center}
    \def\SCALEa{0.3}
    \def\SCALEb{0.3}
    \def\SCALEc{0.3}
    \def\SCALEd{0.3}
    \def\OFFSET{10pt}
    \begin{tabular}{cc}
     \hspace{\OFFSET}
      \includegraphics[scale=\SCALEa]{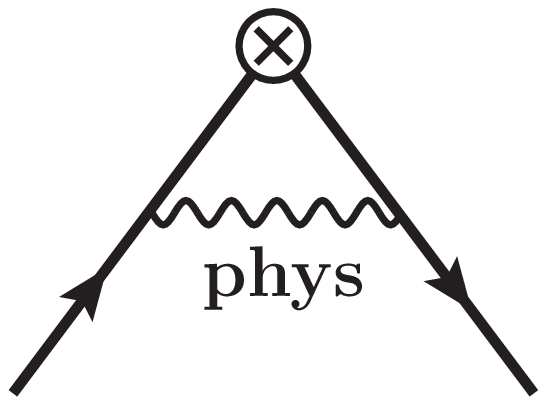} &
     \hspace{\OFFSET}
      \includegraphics[scale=\SCALEb]{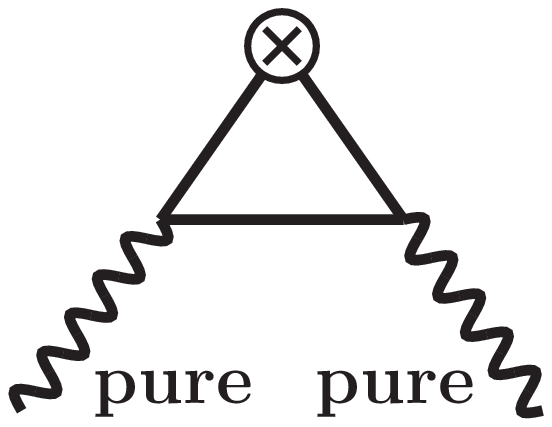} \\
     \hspace{\OFFSET} (a) & \hspace{\OFFSET} (b) \\
      \includegraphics[scale=\SCALEc]{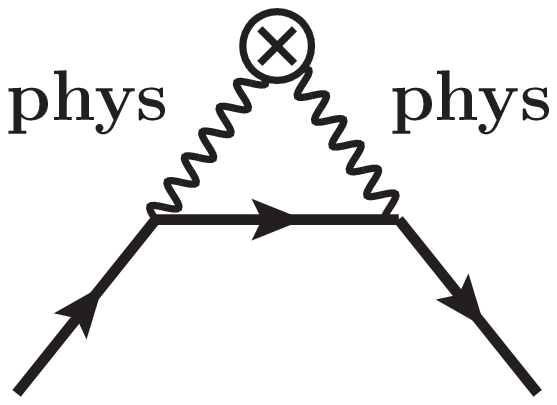} &
     \hspace{\OFFSET}
      \includegraphics[scale=\SCALEd]{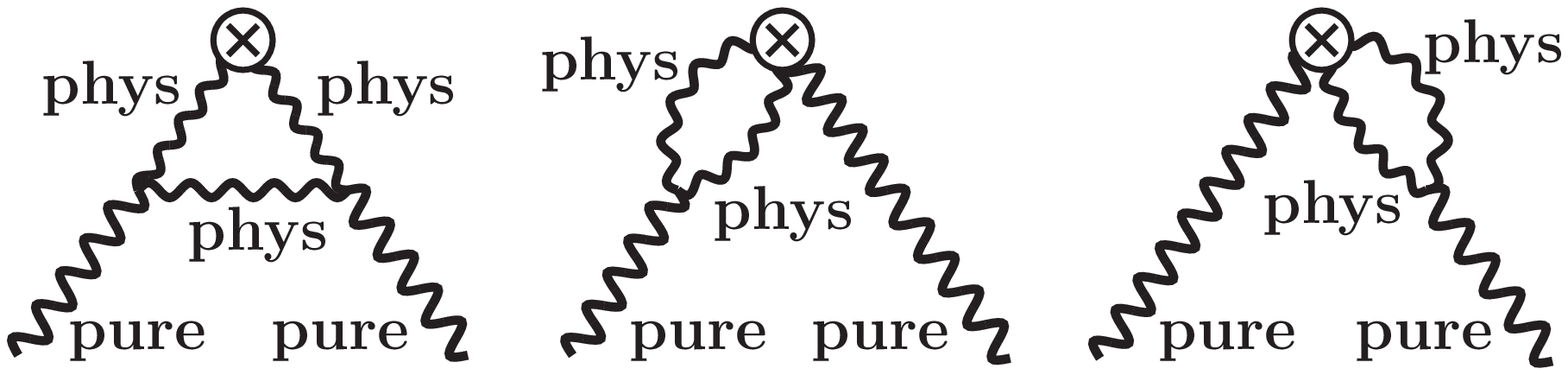} \\
      \hspace{\OFFSET} (c) & \hspace{\OFFSET} (d)
    \end{tabular}
    \caption{One-loop diagrams contributing to the anomalous dimension of the energy-momentum tensor for quarks and gluons in the BFM: (a)~$\gamma^{qq}$, (b)~$\gamma^{qg}$, (c)~$\gamma^{gq}$, and (d)~$\gamma^{gg}$. The symmetric factors to the second and third graphs in (d) are $3/4$ in the BFM; this value can be obtained by calculating $\gamma^{gg}_{\rm\small QCD}$ by the BFM and is different from $1/2$ in the standard method \cite{gamma_QCD}.}
    \label{fig.BFM}
  \end{center}
\end{figure}

Taking into account this principle of the BFM to guarantee the gauge-invariant results, we have to consider Feynman diagrams in Fig.\ref{fig.BFM}~(a),~(b),~(c), and (d).

The fermion lines without arrows should be understood as the sum of clockwise and anticlockwise arrows in Fig.~\ref{fig.BFM}(b) and the contributions from the field renormalization of the quark and the gluon are not explicitly shown. One introduces the so-called {\sl background-gauge-fixing term} and the corresponding ghost-term to the classical QCD Lagrangian in order to covariantlly quantize the theory \cite{Abbott_BFM_Review},
\begin{eqnarray}
\mathcal{L}_{\rm \small gf+gh}
&=& - \frac{1}{\xi}\mbox{Tr}\left(D^{\mu}_{\rm\small pure}A^{\rm\small phys}_{\mu}\right)^2 \nn\\
&{}&- 2 \mbox{Tr}\left( \bar{c}D^{\mu}_{\rm\small pure}D_{\mu}c \right),
\end{eqnarray}
where $\xi$ is the gauge parameter and $(\bar{c})~c$ is the (anti) ghost field. It is notable that the ghost fields do not contribute to the related Feynman-diagrams and only contribute to the field renormalization of the gluon field in external lines. Moreover, we use the dimensional regularization with the dimension $D=4-2\epsilon$ and then the divergence is expressed as a pole of $1/\epsilon$. The anomalous dimensions can be extracted from the coefficients of this pole by differentiating those with respect to the scale $\mu$ introduced in the dimensional regularization. We denote our results of the anomalous dimension based on Eq.~(\ref{eq.gic.Tmunu}) by the BFM as $\gamma_{\rm\small gic}$.

Comparing the diagram of Fig.~\ref{fig.BFM}(a) in our treatment with the standard method (for example, see the related figures in Ref.~\cite{gamma_QCD}, or Ref.~\cite{Wakamatsu_mom_gamma}), the difference between our anylysis with the BFM and the standard method is the absence of (b) of Fig.~2 in Ref.~\cite{Wakamatsu_mom_gamma}. This is obvious because that the gic decomposition of $T^{\mu\nu}_{\rm \small gic,q}$ in Eq.~(\ref{eq.gic.Tmunu}) can only generate one diagram due to the absence of the interaction $\bar{\psi}A_{\rm \small phys}\psi$. It should be emphasized that the definition of $T_{\rm \small gic,q}^{\mu\nu}$ includes $\bar{\psi}A_{\rm \small pure}\psi$ and this $A_{\rm \small pure}$ field cannot generate gluon propagators because of the principle of the BFM. Taking into account this difference, we can recover one of Chen {\sl et al.}'s results, $\gamma^{qq}_{\rm \tiny gic}=\gamma^{qq}_{\rm \tiny Chen}=-2n_g/9$. It is worth noting that the gauge-parameter dependence, $\xi$,  is canceled between Fig.~\ref{fig.BFM}(a) and the field renormalization of the quark within the covariant gauge even in the BFM, as expected. 

This result contradict to Ref.~\cite{Wakamatsu_mom_gamma}. The reason of this disagreement is a misunderstanding of $A_{\rm \small pure}$ and $A_{\rm\small phys}$. The author in Ref.~\cite{Wakamatsu_mom_gamma} indirectly evaluated the contributions to $\gamma_{qq}$ by showing that the so-called potential momentum term vanishes. On the other hand, we can directly derive $\gamma^{qq}_{\rm \small Chen} \neq \gamma^{qq}_{\rm \small QCD}$ thanks to the differences of allowed Feynman-diagrams in the BFM. Similarly, taking into account the BFM, Fig.~\ref{fig.BFM}(c) gives $\gamma^{gq}_{\rm \tiny gic}=\gamma^{gq}_{\rm \tiny Chen}=2n_g/9$ and the reason of the disagreement between ours with Ref.~\cite{Wakamatsu_mom_gamma} is the same reason mentioned in computing $\gamma^{qq}_{\rm \small gic}$, namely, misunderstanding of the internal and external lines. For example, we cannot use Fig.~3 in Ref.~\cite{Wakamatsu_mom_gamma} within the BFM to guarantee the gauge-invariant results.
 
Our results so far seemingly show the consistency between the BFM and Chen {\rm et al}.'s results in the quark sectors $(\gamma^{qq}_{\rm \tiny gic}, \gamma^{gq}_{\rm \tiny gic})^t$. However, Figs.~\ref{fig.BFM}(b) and (d) to calculate the gluonic sectors $(\gamma^{qg}_{\rm \tiny gic},\gamma^{gg}_{\rm \tiny gic})^t$ cause problems of renormalizations in the gluonic sector. We can discuss the divergence of Fig.~\ref{fig.BFM}(b) without any computation in the following way. The BFM does not change the fermion propagator and it gives the same operator-insertion-vertex and there is no difference between the coupling of $\bar{\psi}A_{\rm \small phys}\psi$ and $\bar{\psi}A_{\rm \small phys}\psi$ interaction. Hence Fig.~\ref{fig.BFM}(b) gives the same divergence with the standard computation. Then this divergent-structure seems to give the result, $\gamma^{qg}_{\rm \tiny gic}=\gamma^{qg}_{\rm\small Chen}=\gamma^{qg}_{\rm \tiny QCD}$, and seems to confirm Chen {\sl et al.}'s result at first glance. However, we should carefully check the renormalization of this divergence and the gauge-parameter dependence in the gluonic sectors by the BFM.  
	
As we have already mentioned the principle of the renormalization in the BFM, the divergence should be removed by the counter term defined by the background fields, namely, $A_{\rm \small pure}A_{\rm \small pure}$ term in the definition $T^{\mu\nu}_{\rm\small gic,g}$ in Eq.~(\ref{eq.gic.Tmunu}). However, $T^{\mu\nu}_{\rm \small gic,g}$ in Eq.~(\ref{eq.gic.Tmunu}) does not include such a term and hence we cannot remove this divergence in Fig.~\ref{fig.BFM}(b) due to the lack of the counter term. We can easily check the absence of $A_{\rm \small pure}A_{\rm \small pure}$ term by using the relation, $F^{\mu\nu}=F^{\mu\nu}_{\rm \small pure}+D^{\mu}_{\rm \small pure}A^{\nu}_{\rm \small phys}-D^{\nu}A^{\mu}_{\rm \small phys},$ with the condition $F^{\mu\nu}_{\rm \small pure}=0$. This absence of the counter term means that we cannot correctly define the anomalous dimension in the gluonic sector based on Eq.~(\ref{eq.gic.Tmunu}). It is obvious that we have the same problem in Fig.~\ref{fig.BFM}(d) because of the absence of the counter term. Hence the gic decomposition of the energy-momentum tensor defined in Eq.~(\ref{eq.gic.Tmunu}) causes the inconsistency in the renormalization of the gluonic sectors.

Although we have discussed the gic decomposition of the energy-momentum tensor of quarks and gluons in Eq.~(\ref{eq.gic.Tmunu}), we can easily apply this logic to the gic decomposition of the generalized-angular-momentum tensor for quarks and gluons, for example, as discussed in Refs.~\cite{Leader_Review,Wakamatsu_Review}:
\begin{eqnarray}
 M^{\mu\nu\rho}_{\rm \small gic} 
 &=&    M^{\mu\nu\rho}_{\rm \small gic,q,spin} 
     +  M^{\mu\nu\rho}_{\rm \small gic,g,spin}\nn\\
&{}&     +  M^{\mu\nu\rho}_{\rm \small gic,q,OAM}
     +  M^{\mu\nu\rho}_{\rm \small gic,g,OAM},
\end{eqnarray}
where each term is given by
\begin{eqnarray}
 M^{\mu\nu\rho}_{\rm \small gic,q,spin} 
 &=& \frac{1}{2}\epsilon^{\mu\nu\rho\sigma}\bar{\psi}\gamma_{\sigma}\gamma_{5}\psi,\nn\\
 M^{\mu\nu\rho}_{\rm \small gic,g,spin}
 &=& 2\mbox{Tr}
 \left[   F^{\mu\rho}A^{\nu}_{\rm \small phys} 
        - F^{\mu\nu}A^{\rho}_{\rm \small phys}
 \right],\nn\\
 M^{\mu\nu\rho}_{\rm \small gic,q,OAM}
 &=& \bar{\psi}\gamma^{\mu}
 \left(   x^{\nu}iD^{\rho}_{\rm \small pure}
        - x^{\rho}iD^{\nu}_{\rm \small pure} 
 \right)\psi,\nn\\
   M^{\mu\nu\rho}_{\rm \small gic,g,OAM}
   &=& 2\mbox{Tr}
   \left[   F^{\mu\alpha} 
         \left(   x^{\nu}D^{\rho}_{\rm \small pure} 
                - x^{\rho}D^{\nu}_{\rm \small pure} 
         \right) A_{\alpha,{\rm\small phys}} 
  \right], \label{eq.gic.Mmunurho}\nn\\
\end{eqnarray}
where we ignored the so-called boost term.
Checking the two-point interaction by $A_{\rm \small pure}A_{\rm \small pure}$ in $M^{\mu\nu\rho}_{\rm \small gic,g,OAM}$, we can easily find the absence of such a term in the above definition. Hence we cannot remove the divergences in the gluonic sectors with keeping the principle of the BFM as we discussed the energy-momentum tensor.

It is instructive to try to consider a trick to avoid this problem, that is, the absence of the gluonic counter-term. 
At first, we can recognize that a reason why we do not have $A_{\rm \small pure}A_{\rm \small pure}$ interaction is because of the condition, $F^{\mu\nu}_{\rm \small pure}=0$, at the definitions in Eq.~(\ref{eq.gic.Tmunu}). Hence we may try to keep this term as the nonzero contribution and set it zero at the end of all calculation. Then the gluonic term $T^{\mu\nu}_{\rm \small gic,g}$ is corrected by the new form:
\begin{eqnarray}
 T^{\mu\nu}_{\rm \small gic^{\prime},g}
 &=& - \mbox{Tr}\left[ F^{\{\mu\alpha}(F^{\nu\}}_{\rm\small pure,\alpha}+ D^{\nu\}}_{\rm\small pure}A_{\alpha,{\rm\small phys}}) \right], \label{eq.gic.Tmunu.prime}
\end{eqnarray}
where "${\rm \small gic^{\prime}}~$" stands for an alternative gic-decomposition and the quark part $T^{\mu\nu}_{\rm\small gic, q}$ and the Feynman rule of $A_{\rm\small phys}A_{\rm\small phys}$ interaction is not affected by this modification. Then, we can use $A_{\rm\small pure}A_{\rm\small pure}$ term for the renormalization of the gluonic sector and we reproduce the result $\gamma^{qg}_{\rm\small gic^{\prime}}=\gamma^{qg}_{\rm\small Chen}=\gamma^{qg}_{\rm\small QCD}$. It should be stressed that we do not assume the well-known relation $\gamma^{gg}_{\rm\small QCD}=-\gamma^{qg}_{\rm\small QCD}$ in the standard QCD, which was assumed in Ref.~\cite{Wakamatsu_mom_gamma}; instead, we should directly check whether the gauge-parameter dependence cancels out or not in the nontrivial gluon-to-gluon-sector. Actually this cancellation works in Eq.~(\ref{eq.gamma.QCD}) based on the Eq.~(\ref{eq.Bel.Tmunu}) both by the standard method and by the BFM.

To calculate the last piece, $\gamma^{gg}_{\rm\small gic^{\prime}}$, we have to change the related Feynman-rules of $A_{\rm\small pure}A^2_{\rm\small phys}$ interaction according to the new definition in Eq.~(\ref{eq.gic.Tmunu.prime}) and we use {\sc Package-X} \cite{PackageX} for symbolic calculations. Combining the result of $\gamma^{gg}_{\rm\small gic^{\prime}}$ based on Eq.~(\ref{eq.gic.Tmunu.prime}) with former results, $\gamma^{qq}_{\rm\small gic},\gamma^{gq}_{\rm\small gic}, \gamma^{qg}_{\rm\small gic^{\prime}}$, we denote our final-results as $\gamma_{\rm\small gic^{\prime}}$. The result is summarized in the following:
\begin{eqnarray}
 \gamma_{\rm \tiny gic^{\prime}} 
&=&  - \frac{\alpha_s(\mu)}{4\pi}
\left(
 \begin{array}{cc}
   - \frac{2}{9}n_g &  \frac{4}{3}n_f\\
     \frac{2}{9}n_g  & -\frac{4}{3}n_f - 6 - 2\xi
\end{array}
\right),\label{eq.gamma.Chen2}
\end{eqnarray}
where this result does not coincide with Eqs.~(\ref{eq.gamma.Chen}) or (\ref{eq.gamma.QCD}). Beside, more importantly, the gauge-parameter dependence, $\xi$, remains in the above result of $\gamma^{gg}_{\rm \tiny gic^{\prime}}$ sector and this was not studied in Ref.~\cite{Chen2} because of their explicit-Coulomb-gauge-fixing. This result can be understood as consequences of the modified Feynman-rule for $A_{\rm\small pure}A^2_{\rm\small phys}$ term and of the surface term (potential-momentum term) between Belinfante and gic definition; thus, the cancellation of the gauge parameter working perfectly in the Belinfante definition is spoiled in $\gamma^{gg}_{\rm \small gic^{\prime}}$ due to these effects.  Actually we have already seen the similar effect in $\gamma^{qq}_{\rm\small gic}$ sector, because the surface term changed the loop structure and consequently it gives $\gamma_{\rm \small gic}^{qq} \neq \gamma_{\rm \small QCD}^{qq}$. Hence the above result depends on a choice of gauge and shows that the gic$^{\prime}$ decomposition in Eq.~(\ref{eq.gic.Tmunu.prime}) does not give the gauge-independent result of the anomalous dimension at the one-loop order, even after we adopted the trick to overcome the problem of the gluonic sectors.  

In conclusion, we showed that our analysis of the anomalous dimension matrix for the gic decomposition of the energy-momentum tensor for quarks and gluons by the BFM leads to the inconsistency in the renormalization of the gluonic sectors at the one-loop order. Most importantly, the final result in Eq.~(\ref{eq.gamma.Chen2}) depends on the gauge even after the improvement of this inconsistency. Although both the gic and gik decomposition seems to be possible at the definitions \cite{Leader_Review, Wakamatsu_Review}, our analysis based on the BFM reveals the serious problems in the gluonic sector of gic~(gic$^{\prime}$) decompositions. This result can be extended to the anomalous dimension for the total angular momentum of quark (gluon), $\vec{J}_{q(g)}$,  derived from Eq.~(\ref{eq.gic.Mmunurho}), because the short-distance behavior of $\vec{J}_{q(g)}$ in the standard method is essentially same with that of the quark (gluon) momentum $\vec{P}_{q(g)}$, as shown in Ref.~\cite{Ji_RGE_OAM_qg}. Hence it is obvious that the same problems and inconsistencies appear in the anomalous dimensions of the gic-spin decomposition. We have to check the renormalization and the gauge-dependence in the gluonic sector to confirm this perspective, namely, we should study the anomalous dimension of the gluon-to-gluon sector based on the gic-spin decomposition by the BFM. Such a point of view will shed  light on unknown aspects of the gik and gic decompositions.

Y.~K. thanks Sven Bjarke Gudnason for useful comments on the background field method, and thanks Jarah Evslin and Emilio Ciuffoli for helpful discussions about operator mixing. This work is supported in part by the National Natural Science Foundation of China under Grant No.~11575254.


\end{document}